\def\spose#1{\hbox to 0pt{#1\hss}}
\def\approxlt{\mathrel{\spose{\lower 3pt\hbox{$\sim$}}
        \raise 2.0pt\hbox{$<$}}}
\def\approxgt{\mathrel{\spose{\lower 3pt\hbox{$\sim$}}
        \raise 2.0pt\hbox{$>$}}}
\def\approxpropto{\mathrel{\spose{\lower 3pt\hbox{$\sim$}}
        \raise 2.0pt\hbox{$\propto$}}}
\mathchardef\twiddle="2218

\def\multleft#1{\hbox to size{\vbox {\halign {\lft{##}\cr #1}}\hfill}\par}
\def\multright#1{\hbox to size{\vbox {\halign {\rt{##}\cr #1}}\hfill}\par}
\def\Mdot{\hbox{$\dot M$}}
\def\mdot{\hbox{$\dot m$}}
\def\<{\thinspace}
\def\cm{{\rm\thinspace cm}}
\def\erg{{\rm\thinspace erg}}

\def\g{{\rm\thinspace g}}

\def\G{{\rm\thinspace G}}
\def\ga{{\rm\thinspace gauss}}

\def\K{{\rm\thinspace K}}

\def\Msun{\hbox{$\rm\thinspace M_{\odot}$}}

\def\s{{\rm\thinspace s}}

\def\cmsq{\hbox{$\cm^2\,$}}

\def\ergpcmq{\hbox{$\erg\cm^{-3},$}}
\def\ergpcmps{\hbox{$\erg\cm^{-3}\s^{-1}\,$}}

\def\ergps{\hbox{$\erg\s^{-1}\,$}}

\def\gpcm{\hbox{$\g\cm^{-3}\,$}}

\def\ps{\hbox{$\s^{-1}\,$}}



\newcommand\beq{\begin{equation}}
\newcommand\eeq{\end{equation}}
\newcommand\beqa{\begin{eqnarray}}
\newcommand\eeqa{\end{eqnarray}}

\documentstyle[11pt,aaspp4]{article}

\begin{document}
\title{Neutrino Trapping and accretion models for Gamma-ray Bursts}
%\title{Eddington limited accretion and Gamma-Ray Bursts} 
\author{Tiziana Di Matteo\altaffilmark{1,2}, Rosalba Perna\altaffilmark{3,4},
Ramesh Narayan \altaffilmark{3,6}}  

\altaffiltext{1}{Max-Planck-Institute f{\" u}r Astrophysik, Karl-Schwarzschild-Str. 1, 85740 Garching bei M{\" u}nchen, Germany}
\altaffiltext{2}{Carnegie-Mellon University, Dept. of Physics, 5000 Forbes Ave., Pittsburgh 15231}  
\altaffiltext{3}{Harvard-Smithsonian Center for Astrophysics, 60
 Garden St., Cambridge, MA 02138} 
\altaffiltext{4}{Harvard Society of Fellows, 78 Mt Auburn Street, Cambridge, 
MA 02138} 
\altaffiltext{6}{Institute for Advanced Study, School of Natural Sciences, Einstein Drive, Princeton, NJ 08540}

\begin{abstract}
Many models of Gamma Ray Bursts (GRBs) invoke a central engine
consisting of a black hole of a few solar masses accreting matter from
a disk at a rate of a fraction to a few solar masses per
second. Popham et al. and Narayan et al. have shown that, for $\Mdot
\approxgt 0.1 \Msun$s$^{-1}$, accretion proceeds via neutrino cooling
and neutrinos can carry away a significant amount of energy from the
inner regions of the disks. We improve on these calculations by
including a simple prescription for neutrino transfer and neutrino
opacities in such regions. We find that the flows become optically
thick to neutrinos inside a radius $R \sim 6-40 R_{\rm s}$ for $\Mdot$
in the range of $0.1 -10 \Msun$s$^{-1}$, where $R_{\rm s}$ is the
black hole Schwarzchild radius. Most of the neutrino emission comes
from outside this region and, the neutrino luminosity stays roughly
constant at a value $L_{\nu} \sim 10^{53} \ergps$. We show
that, for $\Mdot \approxgt 1\Msun$s$^{-1}$, neutrinos are sufficiently
trapped that energy advection becomes the dominant cooling mechanism
in the flow. These results imply that $\nu\bar{\nu}$ annihilation in
hyperaccreting black holes is an inefficient mechanism for liberating
large amounts of energy. Extraction of rotational energy by magnetic
processes remains the most viable mechanism.
\end{abstract}


\section{Introduction}
Most popular models of Gamma Ray Bursts (GRBs) invoke a binary merger
or a collapse involving compact objects. In particular, these include
mergers of double neutron star binaries (Eichler et al. 1989; Narayan,
Paczynski \& Piran 1992; Ruffert \& Janka 1999), mergers of a neutron
star with a black hole (Paczynski 1991; Narayan et al. 1992; Ruffert
\& Janka 2001 and references therein), of helium star with a black
hole (Fryer \& Woosley 1998), ``collapsars'' or ``failed supernovae''
(Woosley 1993; Paczynski, 1998; MacFadyen \& Woosley 1999; MacFadyen,
Woosley \& Heger 2001) and ``supranovae'' (Vietri \& Stella 1998).
All of the above scenarios lead to the formation of a black hole with
a debris torus or disk around it. (The only exceptions are models in
which the GRB energy is provided by the magnetic and rotational energy
of the newly formed neutron star; e.g.; Usov 1992). In order to
understand how the extraordinary amount of energy characteristic of
GRBs can be extracted, we are motivated to further examine the
properties of such compact and massive disks around black holes.

Accretion models in the context of GRBs have been recently discussed
by Popham, Woosley \& Fryer (1999; hereafter PWF), Narayan, Piran \&
Kumar (2001; hereafter NPK) and Kohri \& Mineshige (2002). The
typical mass accretion rates in GRB models are extremely high, of the
order of a fraction of solar mass up to a few solar masses per second.
Under such conditions, the gas photon opacities are also very high and
radiation becomes trapped (see e.g.; Katz 1977; Begelman 1978;
Abramowicz et al. 1988). However, at sufficiently high mass accretion
rates, although energy advection remains important in the outer parts,
the disk becomes dense and hot enough in the inner regions to cool
via neutrino emission. For this reason PWF named these disks
neutrino-dominated accretion flows (NDAFs). This regime is of
particular interest, because neutrinos can, in principle (see
e.g. NPK; Ruffert \& Janka 1999), tap the thermal energy of the disk
produced by viscous dissipation and liberate large amounts of its
binding energy (via the $\nu\bar\nu \rightarrow e^+e^-$ process in regions
of low baryon density).  For this mechanism to be efficient, though,
the neutrinos must escape before being advected into the black hole.

In this paper we investigate the effects of neutrino transport within
the context of NDAFs. By using a simple prescription to account for
neutrino scattering and absorptive opacities, we find that, for
accretion rates $\Mdot \approxgt 1 \Msun \ps$, the gas becomes
increasingly more opaque and neutrinos become trapped. As a result,
energy advection becomes the dominant cooling mechanism in the inner
regions of the flows. We show that, as $\Mdot$ increases, the disk
emitting surface moves further out in radius and the neutrino
luminosity of the flow remains nearly constant. We show that the
accretion flow luminosity plateaus as it approaches the neutrino
Eddington limit of the inner disk and as the neutrino cooling
efficiency decreases.

NPK and PWF also noted the importance of neutrino opacity at $\Mdot
\approxgt 1 \Msun$s$^{-1}$  but did not allow
for neutrino transport effects in their models. Our work therefore
complements the earlier studies of NDAFs. A detailed treatment of
neutrino transfer was included in the numerical simulations carried
out by Ruffert \& Janka (1999), where accretion tori are formed as a
result of neutron star merging. Consistent with our findings, these
authors also show that opacities can be high in such tori. However,
their results are specific to the parameter space covered by the
merger model.

In \S 2 we identify the dominant neutrino opacity sources and outline
the basic equations we use and the approximations we make to the
neutrino transfer problem. In \S 3 we present our numerical results.
We delineate the regions of parameter space in accretion rate and
radius where the flow becomes optically thick to neutrino emission.
In \S 4 we discuss the stability of the accretion flow.  Finally, in \S 5, we
compute the flow luminosity of our model, compare it with the derived
neutrino Eddington limit and discuss various implications of our
results for GRB energetics.

\section{The physical model}

\subsection{Neutrino emission and photodisintegration}
PWF and NPK have calculated models corresponding to steady-state
accretion around a black hole with a high mass accretion rate.  Their
work showed that for the full range of accretion rates of interest
($0.01 - 10\Msun$s$^{-1}$ ), and outside a certain radius $R \sim 100
R_{\rm s}$ (where $R_{\rm s} = 2GM/c^2 = 8.85 \times 10^{5} m_3$ cm
is the Schwarzchild radius of a relativistic compact object of mass
$M=3m_3\Msun$), the disk is advection dominated. Inside this radius,
and for $\Mdot \approxgt 0.1 \Msun$s$^{-1}$, the temperature and density of
the gas become high enough that neutrino cooling takes over fairly
abruptly.

In such accretion flows neutrinos are generated both by neutronization
and by thermal emission. The most significant thermal processes are:

(1) Electron-positron pair annihilation ($e^- + e^+ \longrightarrow
\nu_{i} + \bar{\nu}_{i}$ where $i$ represents both electron-type
neutrinos, $\nu_e$-$\bar{\nu}_e$, and heavy-lepton neutrinos
$\nu_{\mu}, \bar{\nu}_{\mu}, \nu_{\tau}, \bar{\nu}_{\tau}$).  The
neutrino cooling rate per unit volume due to electron-positron pair
annihilation is roughly given by (we take the PWF approximation to the Itoh et
al. 1989; 1990 results):
\beq
q_{\nu_i,\bar{\nu}_{i}}^{-} \simeq 5 \times 10^{33} T_{11}^
9\;\;\;\;\ergpcmps
\eeq
where $T_{11} = T/10^{11} \K$. We assume an equilibrium mixture of $e,
\mu$ and $\tau$ neutrinos and antineutrinos\footnote{Thermal
equilibrium is quickly established (in $\Delta t \approxlt$ 1 ms)
between the neutrinos and the matter via weak, neutral current
annihilation of $e^{+}e^{-}$ pairs (e.g.; Bethe, Applegate and Brown
1980; Salpeter \& Shapiro 1981).} so that the total cooling rate, when
all species are included, is simply obtained by multiplying equation
(1) by $N_{\nu} = 3$, the number of neutrino flavors.

(2) Nucleon-nucleon bremsstrahlung ($n + n \longrightarrow n + n +
\nu_{i} + \bar{\nu}_{i}$). Bremsstrahlung cooling 
is represented by (Hannestad \& Raffelt 1998; Kohri \& Mineshige 2002):
\beq
q_{\rm brem}^{-} \simeq 10^{27} \rho_{10}^{2} T_{11}^5.5\;\;\;\;\;\;\ergpcmps,
\eeq
where $\rho_{10}$ is the scaled density, $\rho_{10} = \rho/10^{10}
\gpcm$.

(3) Plasmon decay. This is the decay rate of the transverse plasmons,
which are normal photons interacting with the electron gas through
$\tilde{\gamma} \longrightarrow \nu_{e} + \bar{\nu}_{e}$, and is estimated
by Ruffert, Janka \& Sch\"{a}fer (1996) to be
\beq
q_{\rm plasmon}^{-} \simeq 1.5 \times 10^{32} T_{11}^9 \gamma_{p}^6
\exp^{-\gamma_p} (1 + \gamma_{p}) \left(2 +
\frac{\gamma_p^2}{1+\gamma_p}\right)\;\;\;\;\;\;\ergpcmps,
\eeq
where $\gamma_{p} = 5.565 \times 10^{-2} \sqrt{(\pi^2 +
\eta_{e}^2)/3}$ and $\eta_{e} = \mu_{e} / kT$, and we solve for 
$\mu_{e}$, the electron chemical potential, using equation (24) in
Kohri \& Mineshige (2002).

The second type of neutrino cooling is due to neutronization reactions.
The most significant of these in this context is electron-positron
pair capture on nuclei ($p + e^- \longrightarrow n + \nu_{e}; \;\;\; n
+ e^{+} \longrightarrow \bar{\nu}_{e} + p$), also know as the URCA process.
The cooling rate per unit volume is given by (see also PWF):
\beq
q_{eN}^{-} = q_{e^{-}p}^{-} + q_{e^{+} n}^{-}
\simeq 9.0 \times 10^{33} \rho_{10} T_{11}^6 X_{\rm nuc}
\;\;\;\;\ergpcmps,
\eeq
where $X_{\rm nuc}$ is the mass fraction of free nucleons
approximately given by (e.g.; PWF; Qian \& Woosley 1996):
\beq
X_{\rm nuc} \approx 34.8 \rho_{10}^{-3/4} T_{11}^{9/8} \exp(-0.61/T_{11}),
\eeq
with an upper bound of unity. This prescription takes into account
the transition from nucleon to $\alpha$-particles which effectively
shuts off the URCA process.  $X_{\rm nuc}$ is typically very close to
zero in the outer disk but photodisintegration breaks down $\alpha$
particles in neutrons and protons once $T$ reaches about $10^{10}$ K.
The photodisintegration process is also responsible for cooling the
gas according to (PWF):
\beq
q_{\rm photo}^{-} \simeq 10^{29} \rho_{10} v \frac{dX}{dr}\;\;\;\;\ergpcmps,
\eeq
where $v$ is the disk radial velocity defined in \S 2.3. 

At low optical depths the emission of neutrinos can be computed
directly from the rates given in equations (1), (2), (3) and (4). NPK
and PWF take into account neutrino cooling as described in equations
(1) and (4) and follow the optically-thin approach in their
computations.

\subsection{Neutrino opacities}
Each neutrino emission process, equations (1)-(4), has an inverse
process corresponding to absorption. In addition, scattering impedes
the free escape of neutrinos from the disk.  The optical depths for
neutrinos are therefore given by the inverse of the processes listed
above plus a contribution from neutral-current scatterings off
nucleons.

The inverse process to equation (1), i.e. the interaction of neutrinos
with one another, gives rise to an optical depth:
\beq
\tau_{a, \nu_{i} \bar{\nu}_{i}} \approx \frac{q_{\nu_i,\bar{\nu}_i}^{-} H} 
{4 (7/8) \sigma T^4} \approx 2.5 \times 10^{-7} T_{11}^5 H,
\eeq
where the scale height $H$ is defined in \S 2.3. The term $(7/8) \sigma
T^4$ is the Fermi-Dirac blackbody luminosity and is the same for each
neutrino flavor, $\nu_{i}$\footnote{Note that there is a factor 1/2
difference between electron-positron pairs and the
neutrino-antineutrino contribution, which comes from comparing the
electron spin degeneracy (2 spin states) to the neutrino spin
degeneracy (one helicity state).}; $\sigma$ is the Stefan-Boltzmann
constant.

The inverse process to equation (4), absorption onto protons or onto
neutrons, leads to a neutrino absorptive optical depth in the flow
given by (see also NPK),
\beq
\tau_{a, eN} \approx \frac{q_{eN}^{-} H} {4 (7/8) \sigma T^4} 
\approx 4.5 \times 10^{-7} T_{11}^2 X_{\rm nuc} \rho_{10} H\;. 
\eeq 

Similarly, the inverse of bremsstrahlung and plasmon processes (given
by Eqs. (2) and (3)) lead to two additional sources of opacity,
$\tau_{a, \rm{brem}}, \tau_{a, {\rm plasmon}}$, which can be worked out
in the same way as for the previous two. In accordance with NPK and
PWF, we find that electron-positron pair annihilation and electron pair
capture on nuclei are always the dominant emission mechanism.
Absorption via the inverse of equations (2) and (3) gives rise to much
smaller absorptive optical depths than from equations (7) and
(8). Although all absorption opacity sources are included in our code,
we will mainly discuss those derived above.

A far more important opacity source (particularly for heavy-lepton
neutrinos) comes from neutral-scattering off nucleons ($\nu_{i} + 
\{n,p\} \longrightarrow \nu_{i} + \{n,p\}$). 
The cross section for momentum transfer for neutrino-nucleon
scattering is given by
\beq
\sigma_{s, N} = C_{s,N} \sigma_{0} \left(\frac{E_{\nu}}{m_e c^2}\right)^2,
\eeq
where $E$ is the center of mass neutrino energy, $m_e$ is the electron
rest mass, $c$ the speed of light, $\sigma_{0}= 1.76\times10^{-44}
\cmsq$, and $C_{s,n} = (1 + 5 \alpha^2)/24$ for neutrino-neutron
scattering and $C_{s,p} = [4(C_V -1)^2 + 5\alpha^2]/24$ for
neutrino-proton scattering, with $C_V = 1/2 + 2\sin^2 \theta_{W}$,
$\sin^2 \theta_{W} \approx 0.23$ and $\alpha = 1.25$ (e.g.; Shapiro \&
Teukolsky 1983).
We calculate the Rosseland mean opacity $\kappa_{s, N}$, by
averaging the neutrino energy over the Fermi-Dirac distribution
function with zero chemical potential. Using
\beq
\left\langle \frac{1}{E_{\nu}^2} 
\right\rangle = \frac{5/(7\pi^2)}{(kT)^2} = \frac{1} {13.8 (kT)^2},
\eeq
where $k$ is the gas Boltzmann constant, we find that the total
opacity due to neutrino scattering is $\kappa_{s, N} = \kappa_{s, n} +
\kappa_{s, p} = 13.8 (C_{s,p} Y_{p} + C_{s,n} Y_{n})\sigma_{0} \rho (k T/m_e
c^2)^2$, where $Y_{n}$ and $Y_{p}$ are the fractions of free neutrons
$n$ and protons $p$, respectively.  For completely dissociated matter,
where this process is relevant, the nucleon mass fractions are $Y_{n}
= 1-Y_{e}$ and $Y_p = Y_e$. Here, we take $Y_{e} \sim 0.5$.  This
leads to a total scattering optical depth given by
\beq
\tau_{s, \nu_{i}} = \rho \kappa_{s, N} H = 2.7 \times 10^{-7}
T_{11}^2\rho_{10} H,
\eeq
which has the same temperature and density dependency as equation (8)
(when $X_{\rm nuc} =1$). Both PWF and NPK found that, for accretion
rates above $\mdot \approxgt 1$, where $\dot{m} = \dot{M} /(\Msun {\rm
s}^{-1})$, the disk becomes optically thick to its own neutrino
emission.  This typically occurs within the inner regions of the
disk. However, neither PWF or NPK  studied the effects of
including, self-consistently in the models, the neutrino opacity
sources described above. Here, we use a simple prescription to include
neutrino transport and discuss its effects on models of NDAFs.

\subsection{Basic assumptions and equations}
In this study, (see also NPK), we adopt basic equations based on
Newtonian dynamics. As in previous work (PWF and NPK), we are
interested in the gross properties of the disks and in particular on
the effects of neutrino transport. For simplicity, we use a steady
state disk model. Even though accretion in a GRB engine is time
dependent, the steady state assumption should be a reasonable
approximation since the viscous timescales in the inner disk, where
all the neutrinos processes become important, are much shorter than
those in the outer disk, where $\Mdot$ is expected to vary.

As in the standard theory of thin accretion disks (e.g., Shakura \&
Sunyaev 1973; Frank, King \& Raine 1992), we consider height-averaged
quantities and write the isothermal sound speed as $c_s^2 = P/\rho$ and
the vertical scale height as $ H = c_s/\Omega_K$, where $\Omega_K =
(GM/R^3)^{1/2} = 2.4 \times 10^4 m_3^{-1} r^{-3/2}
\ps $ is the Keplerian velocity and $r$ is in units of $R_s$. We adopt the
standard Shakura-Sunyaev prescription for the kinematic viscosity
coefficient, $\nu = \alpha c_s^2/\Omega_K$, and scale the
dimensionless parameter $\alpha$ as $\alpha_{-1}=\alpha/0.1$ (PWF,
NPK). The continuity equation and an approximate expression for 
angular momentum balance (e.g., NPK) give the following relation for
the mass accretion rate:
\beq
\dot{M} = 4 \pi R \rho H v \approx  6 \pi \nu \rho H. 
\eeq
Correspondingly, the radial velocity of the gas is $v = 3 \nu/2R $.

In the equation of state we include the contributions from radiation
pressure, gas pressure, degeneracy pressure and neutrino pressure (see
also Popham \& Narayan 1995; PWF and NPK).
\beq
P=\frac{11}{12}aT^4  + \frac{\rho k T}{m_p}
\left(\frac{1 + 3X_{\rm nuc}}{4}\right)+
\frac{2\pi hc}{3}\left(\frac{3}{8\pi m_p}\right)^{4/3} 
\left(\frac{\rho}{\mu_e}\right)^{4/3} + \frac{u_{\nu}}{3},
\eeq
where $a$ is the radiation constant and the factor $11/12$ includes
the contribution of relativistic electron positron pairs (as we expect
the temperature to be significantly above the pair production
threshold). In the degeneracy pressure term, $\mu_e$ is the mass per
electron which we take to be equal to $2$ in agreement with NPK and
PWF.  Neutrinos also contribute to the equation of state and this is
taken into account in the forth term, where $u_{\nu}$ is the neutrino
energy density defined as $u_{\nu} = 7/8 aT^4 \sum (\tau_{\nu_{i}}/2
+ 1/\sqrt{3})/ (\tau_{\nu_{i}}/2 + 1/\sqrt{3} + 1/3\tau_{a,\nu_{i}})$
(Popham \& Narayan 1995; see also eq.(15)).

We write the energy equation as
\beq 
q^{+} = \frac{3GM \dot{M} }{8\pi R^{3}} = q^{-} + q_{\rm adv} 
\eeq
where $q^{+}$ represents the viscous dissipation, $q^{-}= q_{\nu}^{-}
+ q_{\rm photo}^{-} H$ is the total cooling rate due to both neutrino
losses ($q_{\nu}^{-}$) and photodisintegration ($q_{\rm photo}$,
Eq. 6) and $q_{\rm adv}$ is the advective cooling rate. All the $q$'s
correspond to half the disk thickness. For the neutrino flux
$q_{\nu}^{-}$, we use a simplified model for the transport, based on
the two-stream approximation and derived by Popham \& Narayan (1995;
see also Hubeny 1990). We write:
\beq
q_{\nu}^{-} =  \sum_{i} \frac{(7/8) \sigma T^4} {(3/4)(\tau_{\nu_{i}}/2 +
1/\sqrt{3} + 1/3\tau_{a,\nu_{i}})}.
\eeq
Here $\tau_{\nu_{i}} = \tau_{a,\nu_{i}} + \tau_{s,\nu_{i}}$ is the sum
of absorptive and scattering optical depths calculated for each
neutrino flavor and $\tau_{a,\nu_{i}}$ the total absorptive optical
depth for each neutrino flavor (e.g. for $\nu_{e}$, the absorptive
optical depth includes all terms, $\tau_{a, \nu_{e}} =
\tau_{a, \nu_{e} \bar{\nu}_{e}} + \tau_{a, eN} + \tau_{a, \rm{brem}} +
\tau_{a, \rm{plasmon}}$ and the last two terms are negligible).
Equation (15) gives the correct behavior in the limit of both small
and large $\tau_{\nu_{i}}$ and $\tau_{a,\nu_{i}}$\footnote{Equation
(15) was derived in the context of radiative transport and assumes that
opacities and emissivities are independent of the $z$
coordinate. Although we do not solve for the vertical structure of the
flow, this assumption is less accurate for the neutrino transport where
all cross sections are a function of temperature and hence of the
vertical disk structure}.

For simplicity, we approximate $q_{\rm adv}$ by, (see e.g., Narayan
\& Yi 1994; Abramowicz et al. 1995):
\beq
q_{\rm adv} = \Sigma v T \frac{ds}{dr} \simeq \xi  v \frac{H}{R} T
\left( \frac{11}{3} a T^3 + \frac{3}{2} \frac{\rho k}{m_p}
\frac{1+X_{\rm nuc}}{4}\right),
\eeq
where $\xi \propto (-d\ln s/d\ln r)$ (see e.g.; Kato, Fukue \&
Mineshige 1999 for the exact expression for $\xi$) is taken to be
constant and equal to $1$.  The factor $(11/3) a T^{3}$ is the entropy
density of the radiation.  The entropy density of neutrinos, a factor
$(7/6) a T^4 = 4/3 \times 7/8 a T^4$, is also added to equation (16)
when the corresponding pressure term is added to the equation of
state. Note that degeneracy pressure makes no contribution to the
advection term because there is no associated entropy.

We solve numerically equations (13)--(16) to find the temperature and
density at a given radius and for a given $\dot{m}$. We are interested
primarily in the properties of the inner regions of the accretion
flow, where neutrino processes are important.  We thus concentrate on
the range of radii from $r_{\rm min}=3$ to $r_{\rm max}= 200$. (From
the work of PWF and NPK we know that, for $r \ga 100$, the flows are
fully advection dominated since neutrino cooling is not important and
photons are completely trapped).

\section{Numerical results}
\subsection{Gas profiles, neutrino optical depths and advection}
Temperature and density profiles calculated from our model are shown
in Figure 1 and the corresponding pressure components in Figure 2.  We
show our solutions for three values of the accretion rate, $\mdot =
0.1,1, 10$ (long dashed, solid and short dashed line,
respectively). The thinner lines show the corresponding PWF solutions
(see also their Figure 3). For easier comparison between the two sets of
solutions, we extend our solution range out to $r=1000$. PWF and NPK
have shown that, at large radii, densities and temperatures are too small
for neutrino cooling to be significant, while optical depths are too
large for photon cooling so that the flows are simply advection dominated.
Therefore the solutions should not be sensitive to any assumption made  
about neutrino cooling and should compare well in this region. Figure 1
shows that, at large radii, there is indeed good agreement, both in the slope
and in the normalization, between our solutions and those of PWF. In this
region, radiation pressure dominates for $\mdot < 10$, whereas the
degeneracy and gas pressure components become increasingly more dominant
for the higher values of $\mdot$ (Figure 2). In the region from $r
\sim 100$ to $\sim 300$, the temperature profiles flatten. Here, 
photodisintegration and, more importantly, neutrino emission starts
cooling the gas. The treatment of neutrino transfer inside this region
becomes important and the two solutions start differing. Note that the
onset of photodisintegration produces a sharp feature in the PWF
solution in this range of radii. Such a feature is also reproduced in
our calculation although it is somewhat weaker, particularly as seen
in the density profiles.
 
Inside radii of $r \sim 30-100$, and for $\mdot \approxgt 1$, our
temperature solution steepens again whereas the PWF profile
flattens. A similar trend is observed in the density profiles. In this
region gas pressure becomes the dominant term in the equation of state
(Figure 2).

Figure 3 shows contour plots for three quantities: the optical depths
$\tau_{a,\nu} = \sum \tau_{a, \nu_{i}}$ (dotted lines) and
$\tau_{s,\nu} = \sum \tau_{s,\nu_{i}}$ (dashed line) in the top
panel, the advection parameter, $f= q_{adv}/q^{+} = 1- q^{-}/q^{+}$
in the middle panel and the viscous timescale, $t_{\rm acc} = R^2/\nu
= \alpha^{-1} (H/R)^{-2}\Omega_{K}^{-1}$ in the bottom panel.
These quantities have been calculated for $\mdot$ ranging from $0.1$
to 10 in a region of the disk from $r=3$ to 200 (where neutrino
cooling is important). The top panel of Figure 3 shows that the
optical depth is dominated by the absorptive opacity in the inner
regions of the flow (and in particular by the inverse of proton and
neutron capture, given by Eq.~4). The optical depth exceeds 1 for $r$
in the range  $\approxlt 5 - 30$ and $\mdot = 0.1 - 10$.  Note that
the cooling function in equation (15) reduces  to the optically
thin expression for small optical depths but it differs significantly
from the latter at optical depths $\sim 1$ and larger. The treatment
of neutrino transport via equation (15), in the optically thick
regions, is what gives rise to the major differences between our
work and that of PWF.

The most significant consequence of the gas becoming opaque is that
neutrinos are trapped in the inner regions of the flow and the
neutrino emission is partially suppressed. This is illustrated by the
middle panel in Figure 3. Energy advection becomes important (i.e. $f
> 0.5$: advection dominates over cooling) in the inner region of the
flows and in particular in the region where the optical depths become
significantly larger than 1. By comparing the contours of the optical
depths and $f$ one finds that the region where most of the neutrino
cooling occurs (where $f$ decreases) lies roughly along (or just
outside) the $\tau =1$ contours. With increasing $\mdot$ the neutrino
cooling becomes increasingly less efficient as the $\tau =1$ surface,
moves further out in the flow. Consistently with previous work, we
also find that $f$ (i.e. advection) also dominates at large radii. In
particular, $f\sim 1$ for $r\approxgt 100$; this is equivalent to the
statement that the cooling timescale is much longer than the accretion
timescales, so the energy is advected inward before it can be radiated
away.

In our solution, which is mostly advection-dominated, we have $H\sim
R$. Therefore, $t_{\rm acc}$ is almost independent of $\mdot$ and
approaches the simple dependence: $t_{\rm acc} \sim
1/\alpha\Omega_{K}$.  Our values of $t_{\rm acc}$ are typically a
factor $\sim 10$ smaller than in Figure 1 of NPK, where the advection
term and the appropriate opacities were not taken into account
self-consistently within the NDAF region.

In the optically thin approximation used by PWF, the neutrino emission
increases with increasing $\mdot$ and energy advection is not
important in the inner regions of the flow.  Although PWF also
estimated that the flows become optically-thick to their neutrino
emission above $\mdot \sim 1$ (consistent with our results), they did
not account for neutrino opacities in their models. Our results,
although based on a simple model, emphasize the importance of
accounting for neutrino transfer\footnote{We note that PWF considered
a fully-relativistic Kerr geometry. Our solution does not take into
account the effects of black hole rotation. However the differences
between our solution and that of PWF are always much larger than those
introduced by allowing for black hole rotation (see e.g.; Figure~6 in
PWF). Although the black hole is most likely rotating, we do not
consider our main conclusions to be significantly limited by adopting
the simple Newtonian treatment. This is because, as shown in Figure 3,
most of the neutrino energy comes from radii of several $R_{\rm
S}$.}. We find that, for most of the range of accretion rates of
interest for GRB models, a neutrino-dominated accretion flow becomes
optically thick to neutrinos, and advection of energy provides the
most significant cooling term in its central regions. In \S 5, we will
discuss the implications of these results for the accretion flow
neutrino luminosities in GRBs.

\section{Stability}
NPK discussed the stability properties of their NDAF solution.  Since
our model differs considerably from theirs, we repeat the analysis
here.

The general condition for thermal stability is:
\beq
\left(\frac{d \ln Q^{+}}{d \ln T}\right)_{|\Sigma} <\left(\frac{d \ln Q^{-}}{d \ln T}
\right)_{|\Sigma},
\eeq
where $Q^{-} = q^{-} + q_{\rm adv}$.  The top panel of Figure 4 shows the
two functions $Q^{+}$ and $Q^{-}$ (solid and dashed line respectively)
as a function of the temperature of the gas.  The radius is fixed at
$r=100$ (for the lower set of curves) and $r=10$ (for the upper set of
curves which have also been renormalized by a factor $10^5$), while
the surface density is taken to be $\Sigma = 10^{16}\gpcm$.  The
equilibrium value of $T$ is given by the intersection of the two
curves.  We see that the stability condition above is satisfied at
this point, so that the solution is thermally stable.

The condition for viscous stability is given by
\beq
\frac{d\dot{m}}{d\Sigma} > 0.
\eeq
The bottom panel of Figure 4 shows $\dot{m}$ versus $\Sigma$ at $r=100$
(solid line) and at $r=10$ (dotted line). This shows that the disk is
viscously stable. We have checked both the thermal and viscous
stability conditions for a variety of $r$ and $\dot{m}$ and we find
that the flows are stable under all conditions. This is a well known
property of flows in which advection is important.

Finally, we also check for the gravitational stability condition,
namely that the Toomre parameter $Q_T$ should be larger than 1.  For a
Keplerian disk, $Q_T$ is given by:
\beq
Q_{T} = \frac{c_s \kappa}{\pi G \Sigma} = \frac{\Omega_K^2}{\pi G \rho} 
\eeq
where $\kappa$ is the epicyclic frequency. Figure 5 shows the
variation of $Q_T$ with radius for $\dot{m} = 0.1, 1, 3, 10$. We see
that $Q_{T}$ decreases with increasing $r$, so that the flows are most
unstable on the outside (as found also by NPK).  However, only for the
largest $\mdot \sim 10$ and for $r\approxgt 50$ does $Q_T$ go below
unity, signifying gravitational instability. Thus the flows are
gravitationally stable under almost all conditions of interest.

\section{Discussion}
The relevance of accretion processes in the central engine of GRBs was
highlighted by Narayan et al. (1992). PWF carried out the first
detailed study of accretion around black holes with ultra-high
$\dot{m}$ (of order a solar mass per second), and identified a new
mode of accretion which they named neutrino-dominated accretion flow
(NDAF). In a subsequent study, NPK noted that accretion flows with
such high $\dot{m}$ are highly advection dominated at large radii
($r\ga 100$). They suggested that the flow at these outer radii may
take the form of a convection-dominated accretion flow (CDAF; Narayan,
Igumenshchev \& Abramowicz 2000; Quataert \& Gruzinov 2000) or a
related kind of flow (e.g., Blandford \& Begelman 1999). As a result,
only a small fraction of the available mass accretes on the black
hole, the bulk of the gas being ejected from the system. Close to the
black hole ($r\la 100$), however, NPK confirmed PWF's result, that a
cooling-dominated NDAF should be present.

PWF and NPK made the simple assumption that the accreting gas is
optically thin to its own neutrino emission. We have improved on this
by including the effects of neutrino transfer (via a simple
prescription) and the neutrino opacities self-consistently in the
model. With these improvements, we find that the central regions of
NDAFs are typically opaque, so that the neutrinos emitted by the
accreting gas are largely trapped. The optically-thick region extends
out to $r \sim 4-5$ at $\mdot \sim 0.1$ to $r\sim 30-40$ at $\mdot
\sim 10$ (Figure 3). We have shown that, above $\mdot \approxgt 1$, the
majority of the energy liberated by viscous dissipation is advected
with the flow instead of being emitted in the form of neutrinos, and
energy advection is much more important in the inner regions of the
accretion flows than previously realized (Figure~3).

In order to further explore the implications of these results we now
calculate the neutrino luminosity from our model and the fraction of
the binding energy carried out to infinity by neutrinos. We also
derive the neutrino Eddington luminosity for such flows.

Figure 6 shows the neutrino luminosity from the accretion flow,
$L_{\nu} = \int_{r_{\rm min}}^{r_{\rm max}} 2 \pi q_{\nu}^{-}rdr$, (in
units of $10^{51} \ergps$; solid line; upper panel). Also, using the
opacities discussed in \S 2.2 we derive the neutrino Eddington
luminosity of the flow:
\beq
L_{\rm Edd,\nu} = \frac{4\pi GM c}{\kappa_{\nu}} \sim 9
\times 10^{53} m_3 T_{11}^{-2} \ergps ,
\eeq
where, for illustration the numerical value given on the right uses
only the two dominant components of the opacity: neutron and proton
absorption and scattering (for electron neutrinos this gives
$\kappa_{\nu_e} =
\kappa_{\nu_e, a} + \kappa_{\nu_e, s} = 5.5 \times 10^{-17} T_{11}^2
\cmsq$g$^{-1}$; Equations 8 and 11; note that for the calculation shown in 
Figure 6 all opacity terms are used). Because the neutrino cross
sections are a function of neutrino energies, the neutrino Eddington
luminosity is a function of both black hole mass and gas temperature.
In Figure 6, we show the neutrino Eddington luminosity calculated for
$T_{11} = T_{11} (r=3)$, the temperature at the inner radius of the
flow; at larger radii the Eddington luminosity increases.  In the
middle panel of Figure 6 we also show the neutrino radiative
efficiency, defined, as in standard accretion theory, by $\eta_{\nu} =
L_{\nu}/\Mdot c^2$.

The neutrino luminosity from the flow is $L_{\nu} \sim 10^{51}
\ergps$ at $\mdot \sim 0.01$  and increases almost linearly 
with $\mdot$ up to $\mdot \sim 0.1$ (Figure 6). Between $0.1 \approxlt
\mdot \approxlt 1$, the luminosity flattens off significantly, ranging
within $6$ -- $8 \times 10^{52} \ergps$. Above $\mdot \sim 1$ the
neutrino luminosity stays virtually constant at $L_{\nu} \sim 8
\times 10^{52} \ergps$. In contrast, in the PWF solution, 
which assumes optically thin neutrino emission, the neutrino
luminosity ranges from $L_{\nu} \sim 10^{51-52} \ergps$ for $\mdot =
0.1$ to up to $L_{\nu} \sim 10^{54} \ergps$ for $\mdot=10$ (see their
Table 3). Thus, one important consequence of the addition of neutrino
opacities is that the flow luminosity varies only by a factor of a few
over the range of accretion rates appropriate for popular models of
GRB progenitors.  The luminosity of the flow is almost constant
because the efficiency with which energy is transported out of the
flow by neutrinos, $\eta_{\nu}$ (bottom panel, Figure 6) decreases
with increasing $\mdot$ (or equivalently, as shown in Figure 5,
decreases as neutrinos become trapped and energy advection becomes
more dominant at larger $\mdot$). We show that $\eta
\sim 0.1$ (consistent with the efficiency expected from a
geometrically thin, cooling dominated Newtonian accretion disk) up to
$\mdot \sim 0.1$ and decreases (almost linearly) with increasing
$\dot{m}$ (reaching $\eta
\sim $a few $ 10^{-3}$ at $\mdot =10$). The fraction of energy
transported away by neutrinos decreases as the inner regions of the
flow become neutrino opaque (see Figure 3).

Figure 6 also shows the neutrino Eddington luminosity evaluated at the
inner edge of the disk (dashed line). In the inner regions of the
flow, the temperature is the highest, and consequently the Eddington
luminosity the lowest (Eq.~20), so that $L_{\nu} \sim L_{{\rm
Edd},\nu}$ for $\mdot \approxgt 1$. Momentum deposition in the inner
regions of the flow by neutrinos may therefore be quite effective as a
mechanism responsible for ejection.

We now examine the implications of our results for GRBs. A number of
authors, including Eichler et al. (1989; see also Narayan et al. 1992)
suggested that $\nu\bar\nu$ annihilation around merging compact
binaries might produce a relativistic $e^{+}e^{-}$ fireball with
sufficient energy to power a GRB. Detailed numerical simulations by
Janka et al. (1999) show that this process might conceivably power
short GRBs (those with durations under a couple of seconds or so),
provided that there is a modest level of beaming. However, we have
shown that neutrino advection is important in most of the parameter
space of these models, which may have a significant impact on the
efficiency of this process. We estimate the luminosity due to
$\nu\bar\nu$ annihilation along $z$-axis above the disk (using an
improved version of Eq.~10 in Ruffert et al. (1997) provided by Thomas
H. Janka, private communication) to be:
\beq
\L_{\nu \bar{\nu}} \sim 6 \times 10^{-35}  \frac{2\;\langle
E_{\nu} \rangle\;L_{\nu}^2}{\pi c^2\left[1-({R_{\rm
min}}/{R_{\rm surf})^2}\right]^2}
\;\;\frac{1}{R_{\rm surf}} \int_{0}^{\infty} d\epsilon\;\Phi\; (\cos
\theta_{\rm min}
- \cos\theta_{\rm surf})^2\;\;\ergps,
\eeq
where the constant in front takes into account the neutrino cross
sections. $L_{\nu}$ is given above and shown in Figure 6
and $R_{\rm surf}$ and $\langle E_{\nu} \rangle $ are evaluated at the
neutrinosphere, defined as the surface at which the emergent neutrinos
originate. Hence, $\langle E_{\nu}
\rangle$ is the average energy of the escaping neutrinos given by
$\langle E_{\nu} \rangle = 3.7\, k T_{11, {\rm surf}}$ (Eq.~10) with $T_{11,
{\rm surf}} = T_{11}(r_{\rm surf}=r(\tau_{\nu}=2/3))$ and $\tau_{\nu} =
\tau_{a,\nu} + \tau_{s,\nu}$.  If the spectrum is a blackbody, the
radius of this surface is located near and above the layer of optical
depth $= 2/3$ (this is because the opacity is mostly dominated by
absorption processes). The contour of $\tau_{\nu} =2/3$ and the radius
of the neutrinosphere as a function of $\mdot$ is shown in Figure
3. We find that the radius of the neutrinosphere \footnote{To derive
the analytical scalings we solve Eqs.~(13) and (14) assuming $P=P_{\rm
gas}$ and $q^{+} = q_{adv}^-$. We have shown that is a good
approximation within the optically thick region.} to be given by,
$r_{\rm surf} = r_{\tau=2/3}
\sim 17 \mdot^{2/5}$ in the range $0.1 \approxlt \mdot \approxlt
10$. At this radius the temperature of the neutrinosphere is roughly
$T_{11,{\rm surf}} \sim 0.2
\mdot^{-3/20}$ which is fairly insensitive to $\mdot$
in the range from $0.1 \approxlt \mdot \approxlt 10$). The integral
part in equation (21) is a geometrical factor, where $\Phi =
3/4[1-2\langle
\mu \rangle^2 +\langle \mu^2 \rangle^2 +1/2 (1-\langle \mu^2
\rangle)^2]$, where $\langle \mu^n\rangle =
\int_{\mu_{\rm surf}}^{\mu_{\rm min}} 
d\mu \mu^n / \int_{\mu_{\rm surf}}^{\mu_{\rm min}} d\mu$ with
$\mu_{\rm surf} =
\cos\theta_{\rm surf} = 1/ \sqrt{1 + (r_{\rm surf}^2/z^2)}$, $\mu_{\rm
min} =
\cos\theta_{\rm min} = 1/ \sqrt{1 + (r_{\rm min}^2/z^2)}$ and $\epsilon =
z/r_{\rm surf}$. 

The bottom panel of Figure 6 shows the neutrino
annihilation luminosity, $L_{\nu \bar{\nu}}$ in units of $10^{51}
\ergps$.  $L_{\nu \bar{\nu}}$ increases up to 
its maximum value of $\sim 10^{50} \ergps$ at $\mdot \approxgt 1$ and
slightly decreases for larger $\mdot$. Our estimate of
$L_{\nu,\bar{\nu}}$ at $\mdot =1$ agrees well with values estimated in
the simulations of Ruffert \& Janka (1999) and PWF (although our
results are less accurate than those of PWF for $\mdot \approxlt
0.1$). In accordance with PWF we find that energetic events can only
be achieved for $\mdot > 0.1$ (below which the neutrino annihilation
efficiency decreases very sharply; see also their Table 3); but in
contrast with their results we do not find that increasingly more
energetic events can be achieved for larger accretion rates. Our
calculations imply that the efficiency of $\nu\bar\nu$ annihilation
remains constant (or slightly decreases) for $\mdot \approxgt 1$. This
is expected, as we have found that neutrinos are increasingly more
trapped in the disk. (Our estimate may be uncertain by a factor $\sim
2$; this is mostly due to the fact that our solution gives only
height-averaged quantities whereas the vertical disk stratification
may be important in this calculation). Note also that recent results from
hydrodynamical calculations carried out by Lee \& Ramirez-Ruiz (2002)
show that $\nu\bar\nu$ annihilation can only produce bursts from
impulsive energy inputs, as the annihilation luminosity scales as
$t^{-5/2}$. This further restricts the importance of this process
to a small fraction of bursts.

Energy extraction from the disk is also possible by MHD processes.
Such processes are broadly based on the expectation that the
differential rotation of the disk will amplify pre--existing magnetic
fields, until they approach equipartition with the gas kinetic energy.
Proposed mechanisms include Parker instabilities in the disk leading
to reconnection, relativistic flares and winds (Narayan et al. 1992;
Meszaros \& Rees 1997) or the Blandford-Znajek mechanism (Blandford
\& Znajek 1977; hereafter BZ). Of these, perhaps, the Blandford-Znajek 
efficiency is the easiest to estimate (at least roughly). We follow
the common assumption that the magnetic field in the disk will rise to
some fraction of its equipartition value $B^2/8 \pi \sim \rho
c^2_s$. Typical values of $\rho c^2_s$ are $10^{30-32} \ergpcmq$ for
$0.1 < \mdot < 10$ (see Figure 2), implying a field strength of $\sim
10^{15 -16}$ G if we make the conservative assumption that $B$ is only
$10 \%$ of its equipartition value. Consistent with earlier work (PWF;
Ruffert \& Janka 1999), the BZ jet luminosity at $\mdot=10$ is then
\beq
L_{BZ} = \left(\frac{B^2}{4\pi}\right) \pi R_{h}^2 a^2 c \approxgt 10^{52}
a^2 \left(\frac{B}{10^{16} \G} \right)^2
\left(\frac{M}{3\;\Msun}\right)^2 \;\;\;
\ergps\;,
\eeq
where $R_{h}= 2 G M/c^2$ is the black hole radius and $a = R_{h}
\Omega_{h}/c$ is the dimensionless black hole spin parameter ($0 <a
<1$). $L_{BZ}$ is about 2 orders of magnitude larger than the neutrino
annihilation rate, $L_{\nu \bar\nu}$, at $\mdot \sim 10$. For larger
equipartition fractions, the ratio $L_{BZ}/L_{\nu \bar\nu}$ becomes
large even at $\mdot \sim 0.1$ or $1$. It is obvious that the energy
liberated by this mechanism is simply proportional to the internal
energy density of the disk and hence is not affected by advection.
Indeed, Livio, Ogilvie \& Pringle (1999) have argued that the BZ
mechanism might be most relevant for advection dominated flows (see
also Meier 2001). Numerical simulations such as those by MacFadyen \&
Woosley (1999) and MacFadyen, Woosley \& Heger (2001) have shown that
the BZ mechanism would be an efficient mechanism capable of liberating
a large fraction of the black hole spin energy.

By similar electromagnetic considerations as those used to derive the
BZ luminosity, Livio et al.(1999) estimate that the
electromagnetic/wind power output from a disk is given by:
\beq
L_{d} = \left(\frac{B_{pd}^2}{4\pi}\right) \pi R_{d}^2 \left(\frac{R_d
\Omega_d}{c}\right) c \sim \left(\frac{B_{pd}}{B}\right)^2 \left(\frac{R_{d}}
{R_{h}}\right)^{3/2} a^{-2} L_{BZ} \sim 5 \; a^{-2} L_{BZ}
\eeq
where, following Livio et al., we take ${R_d \Omega_d}/{c} \sim
(R_h/R_d)^{1/2}$. $B_{pd}$ is the poloidal field in the disk given,
approximately, by $B_{pd} \sim (H/R) B$ (see also Merloni \& Fabian
2002). Because $H/R \sim 1$ in the inner region of the disk we have$
B_{pd} \sim B$. $R_{d}$ is a factor of a few times $R_{h}$ (above we
take $R_{d}$ as the radius of the innermost stable orbit, at which the
disk energy density is the highest). Since $a^2 <1$ and, $R_d$ is at
least a few times $R_{h}$, it is evident that in a geometrically thick
disk, the electromagnetic/wind luminosity from the disk can easily be
larger than the BZ luminosity, hence $L_{d} >> L_{\nu \bar{\nu}}$. Our
calculations therefore indicate that, with increasing accretion rates,
MHD processes become significantly more efficient at releasing energy
than neutrino annihilation processes, hence they are probably the most
viable mechanisms for energy extraction in these systems.

Although the stability analysis presented in \S 4 shows that the
accretion flows we have studied are intrinsically stable, we note that
the viscous time scale (Fig. 3) can be as short as a small fraction of
a second. Therefore, if the accretion disk is fed in a variable
manner, e.g., via fallback material after a supernova explosion as in
the collapsar model, we may expect variations in the accretion rate.
This might explain the complex light curves of GRB in at least some of
the scenarios discussed above (e.g., MacFadyen \& Woosley, 1999).

\acknowledgements
We thank Thomas H. Janka for valuable comments and important
suggestions. We also thank Chris Fryer and Stan Woosley for sending
us the files of their temperature and density solutions used in Figure
1 and Andrea Merloni for discussion.  
T.D.M. aknowledges support by grant NAG-10105.
R.\,P. acknowledges support by the IAS-CNR (Rome, Italy) during the
time that part of this research was conducted.  R.\,N.\ was supported
in part by the W. M. Keck Foundation as a Keck Visiting Professor at
the Institute for Advanced Study.  R.\,N.'s research was supported in
part by NSF grant AST-9820686.

\newpage

{Fig.1 --- Temperature and density profiles in the top and bottom
panel respectively. The profiles are shown for three values of the
accretion rate: $\mdot = 10$ - short dashed lines; $\mdot = 1$ - solid
lines; $\mdot =0.1$ - long dashed lines. The PWF numerical solutions
(thinner lines - see also their Fig. 3) are also plotted for direct
comparison.}

{Fig.2 --- Pressure components for three values of $\mdot$. The gas
	    pressure is shown by the solid line, degeneracy pressure
	    by the dotted line, radiation pressure by the dashed
	    line, neutrino pressure by the long dashed line}

{Fig.3 --- Top panel: Contours showing the values of scattering
optical depth ($\tau_{s}$ - dashed line) and absorptive optical depth
($\tau_{a} $ - dotted line) for a range of $\mdot = 1 - 10$ and $r = 3
- 200$. The $\tau = 2/3$ contour is shown with the solid line.  Middle
panel: Contours showing the advection parameter $f = q_{adv}/q^{+}$
for the same range of $\mdot$ and as a function of $r$. For most
regions of parameter space $f>0.5$ and advection dominates.  Bottom
panel: contours of the accretion time, $t_{acc}$.}

{Fig.4 --- Stability analysis. The top panel shows $Q^{+}$ (dashed
line) and $Q^{-} = q^{-} + q_{adv} $ (solid line) versus $T$ for $r =
100$ (bottom curves) and $r=10$ (top curves - renormalized by a factor
of $10^5$ for better clarity) and for $\Sigma = 10^{16}
\gpcm$. According to the criterion given in Equation 10, the flow is
thermally stable. The lower panel shows $\mdot$ versus $\Sigma$ for $r
=100$ (solid line) and $r=10$ (dotted line). The viscous stability
condition (Eq.~11) is always satisfied.}

{Fig.5 --- The Toomre parameter $Q_{T}$ (Eq. 12) as a function of
radius for the four values of the accretion rate as in Fig. 2. $Q_{T}$
is always $> 1$ (for $\mdot \approxgt 3$) implying that the flow is
gravitationally stable throughout and unstable in the outer regions
for $\mdot \approxgt 3$.}

{Fig.6 --- Upper panel: the solid line represents the neutrino
luminosity and the dashed lines the neutrino Eddington luminosity as a
function of the accretion rate and for two values of the effective
temperature. Middle panel: the solid line shows the radiative
efficiency of the accretion flows as a function of $\mdot$.  Bottom
panel: $\L_{\nu \bar{\nu}}$ as a function of $\mdot$. The dotted line
below $\mdot = 0.1$ represents the most uncertain (upper limit)
estimates. Better values of $\L_{\nu \bar{\nu}} $ for $\mdot < 0.1$
are given in Table 3 of PWF.

\newpage
\begin{figure}[t]
\centerline{\epsfysize=4in\epsffile{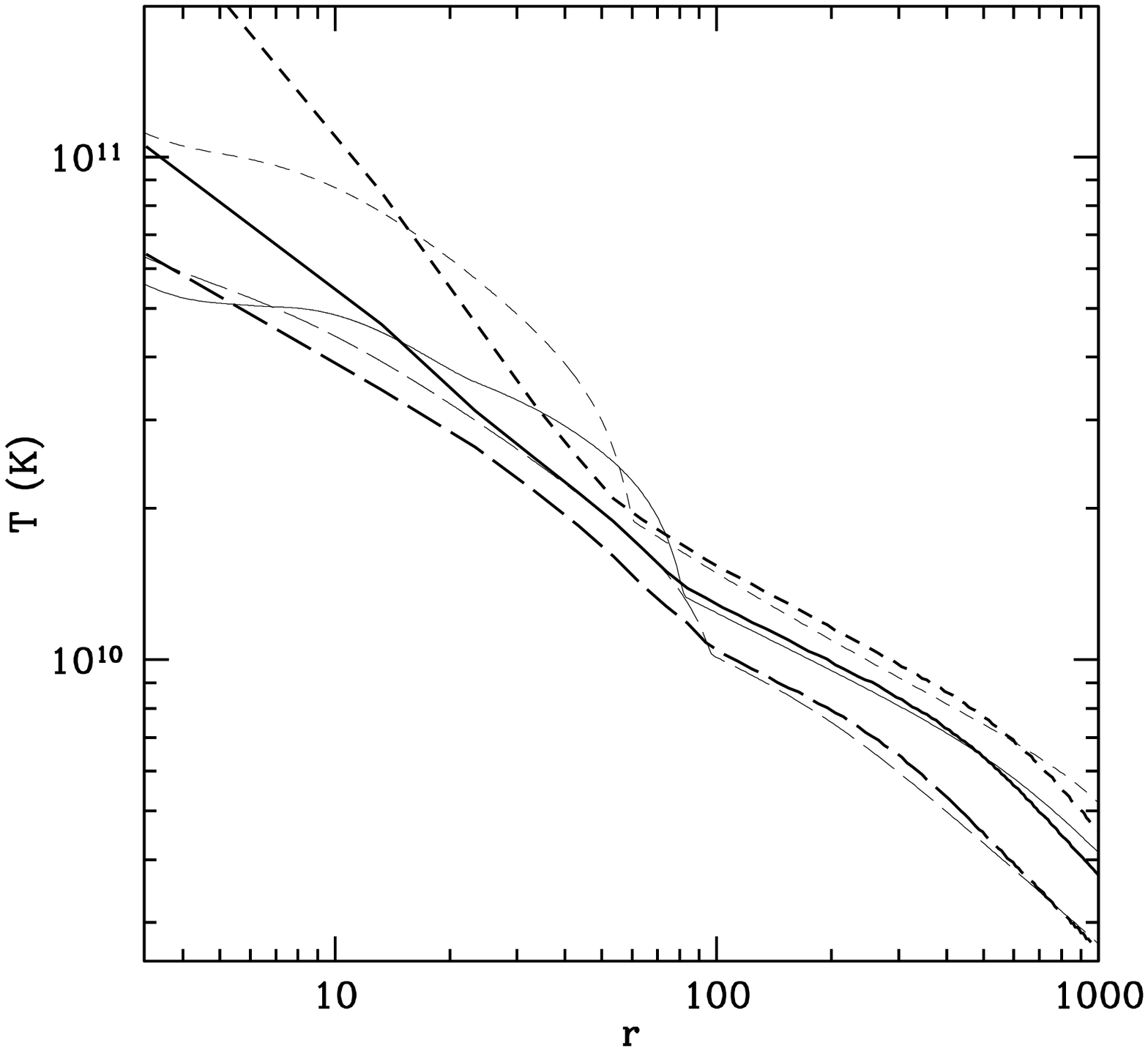}}
\centerline{\epsfysize=4in\epsffile{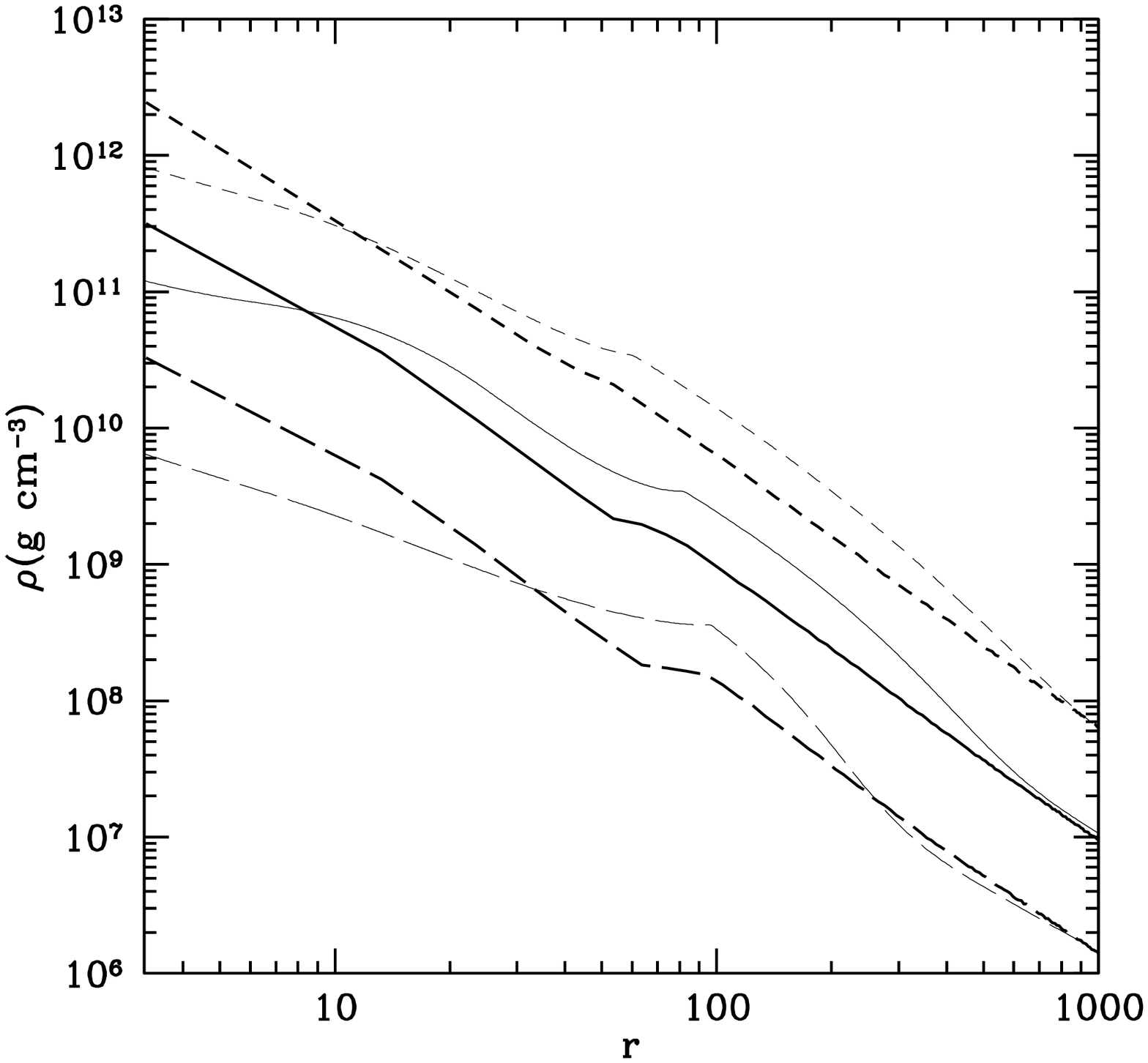}}
\end{figure} 

\begin{figure}[t]
\centerline{\epsfysize=5.7in\epsffile{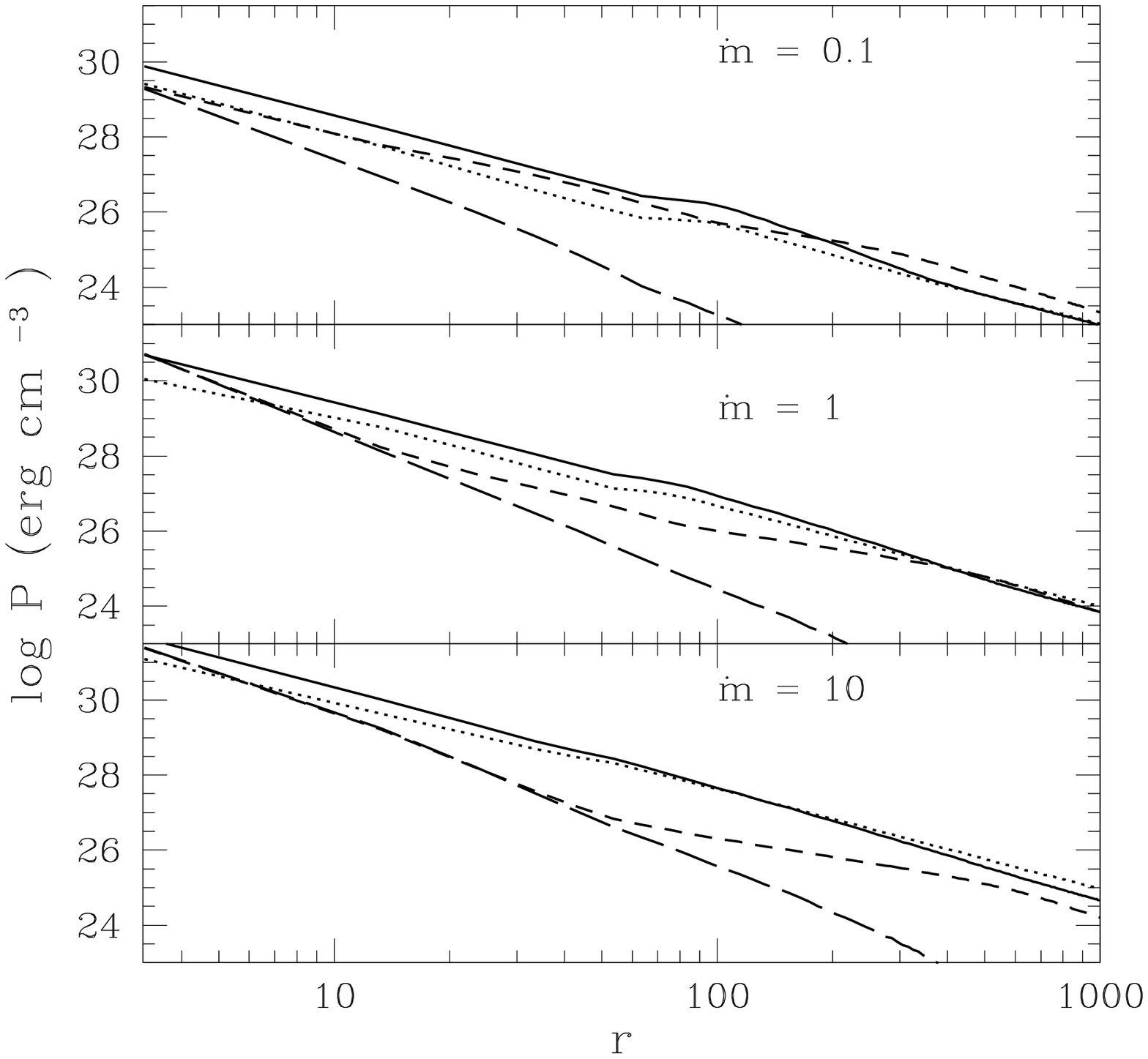}}
\end{figure} 

\begin{figure}[t]
\centerline{\epsfysize=2.7in\epsffile{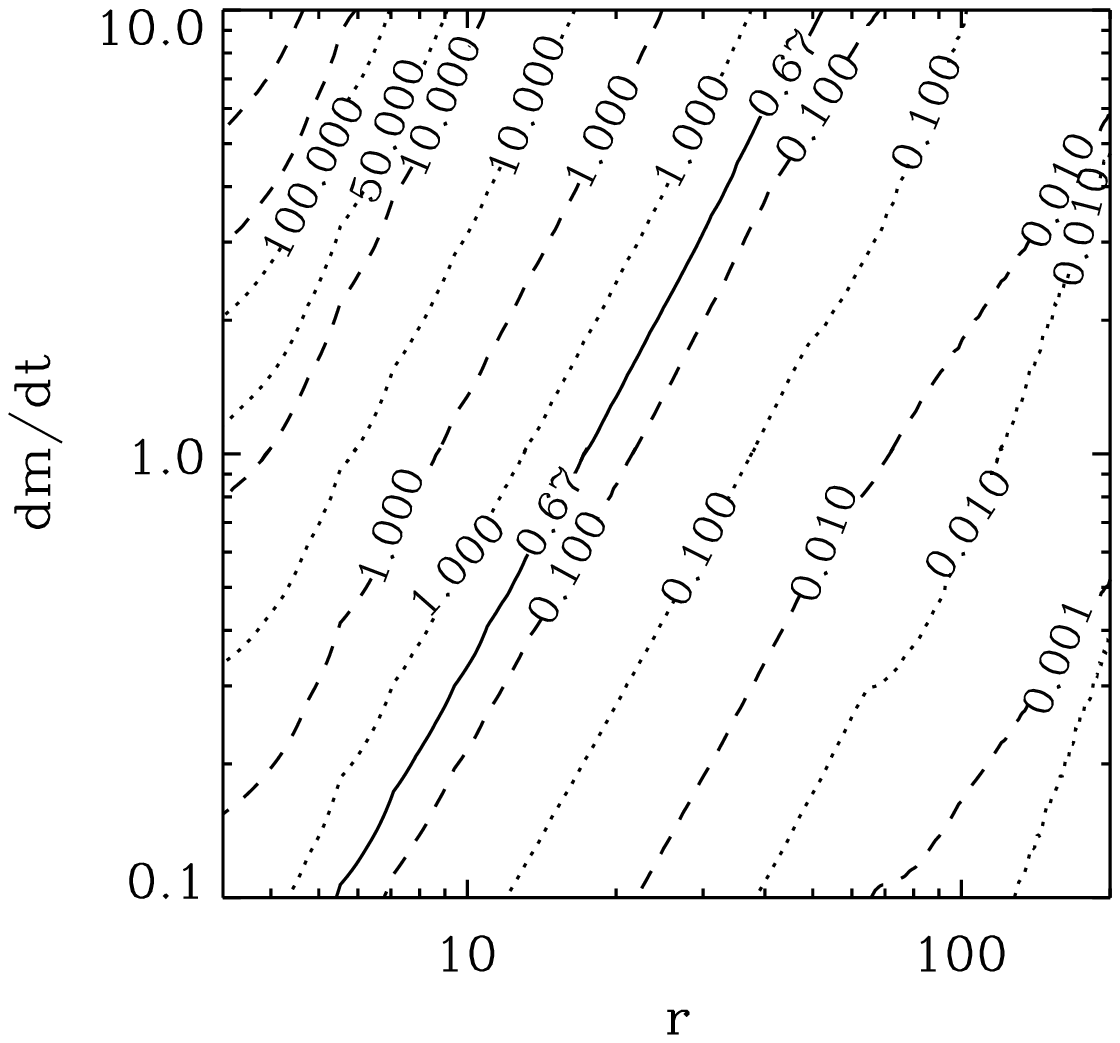}}
\centerline{\epsfysize=2.7in\epsffile{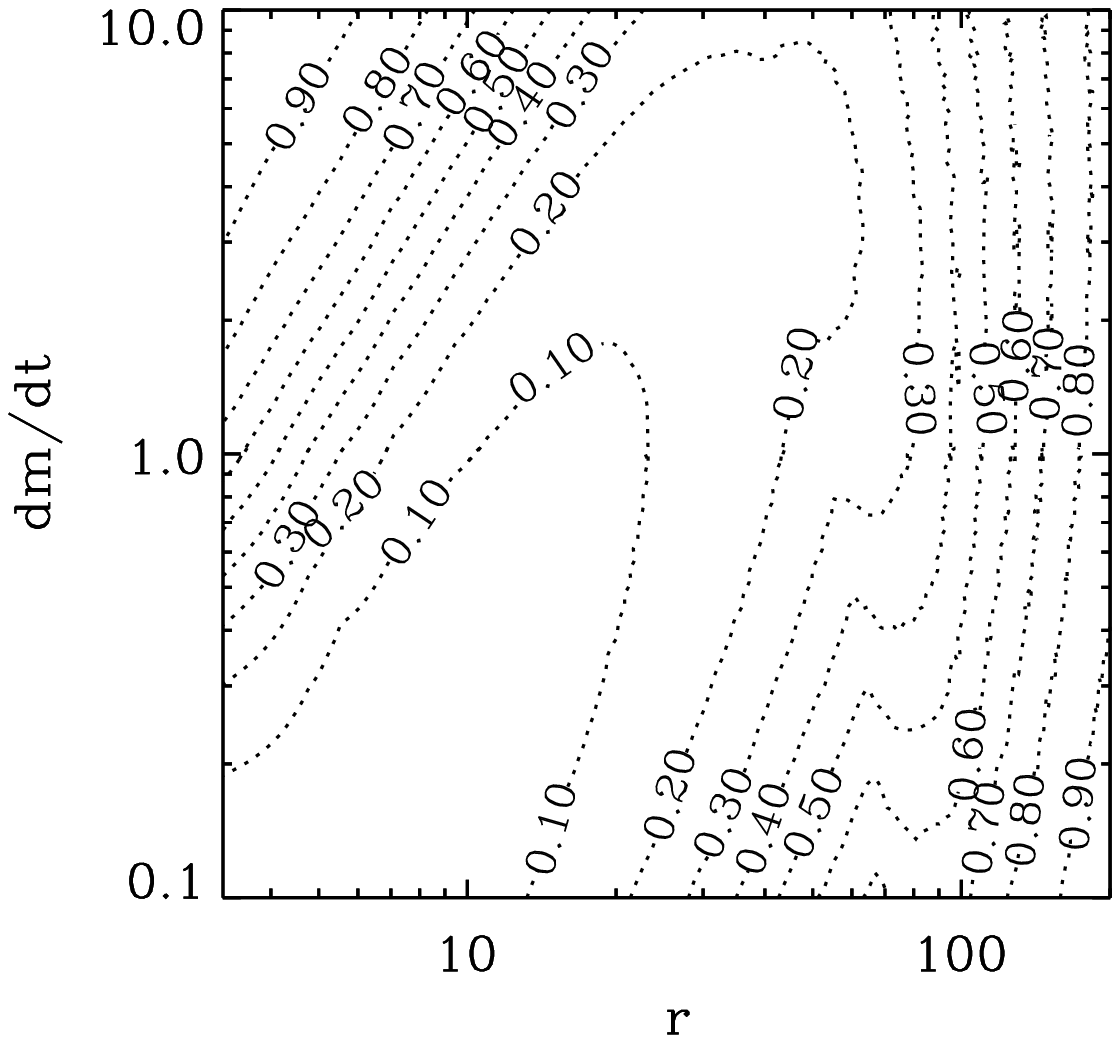}}
\centerline{\epsfysize=2.7in\epsffile{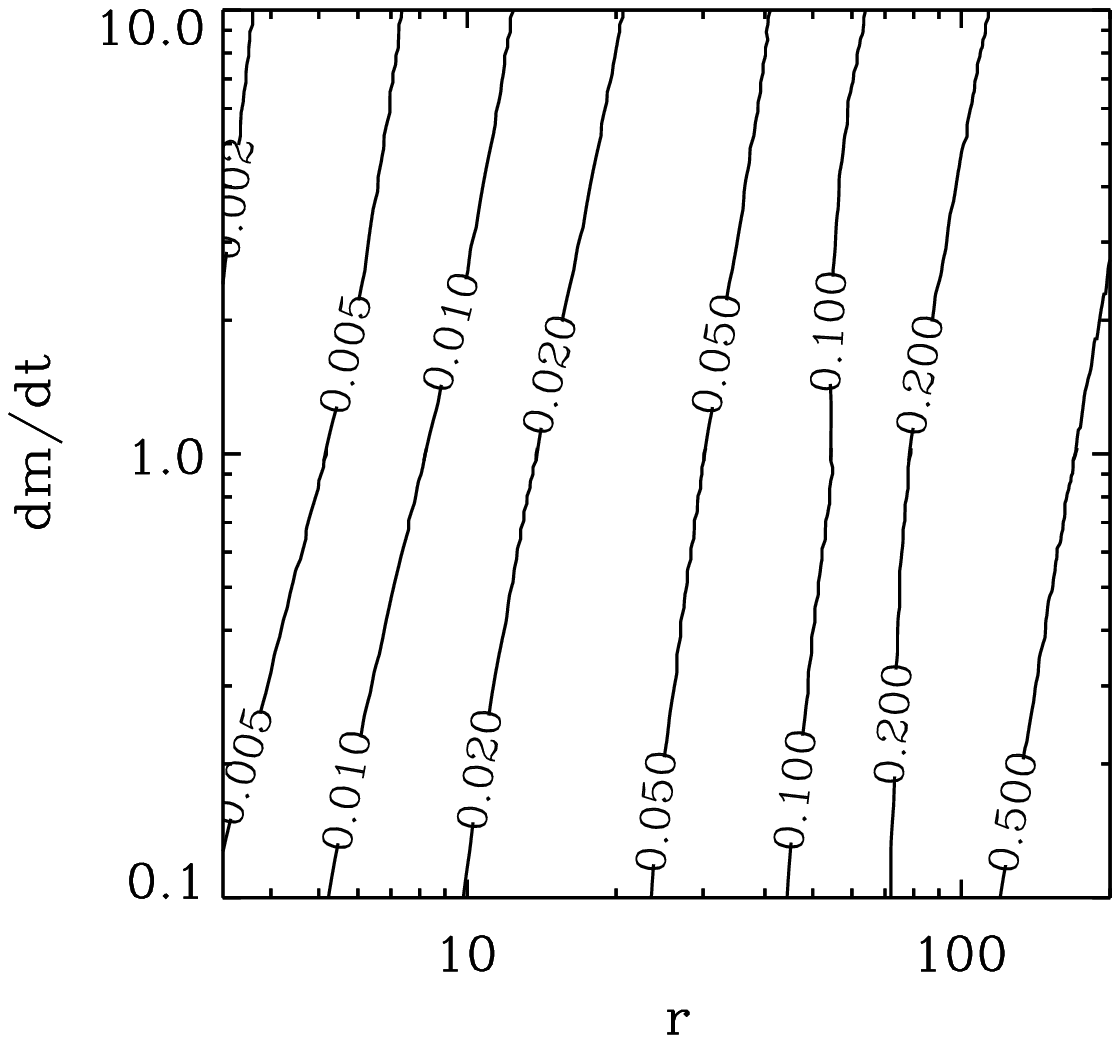}}
\end{figure} 

\begin{figure}[t]
\centerline{\epsfysize=4in\epsffile{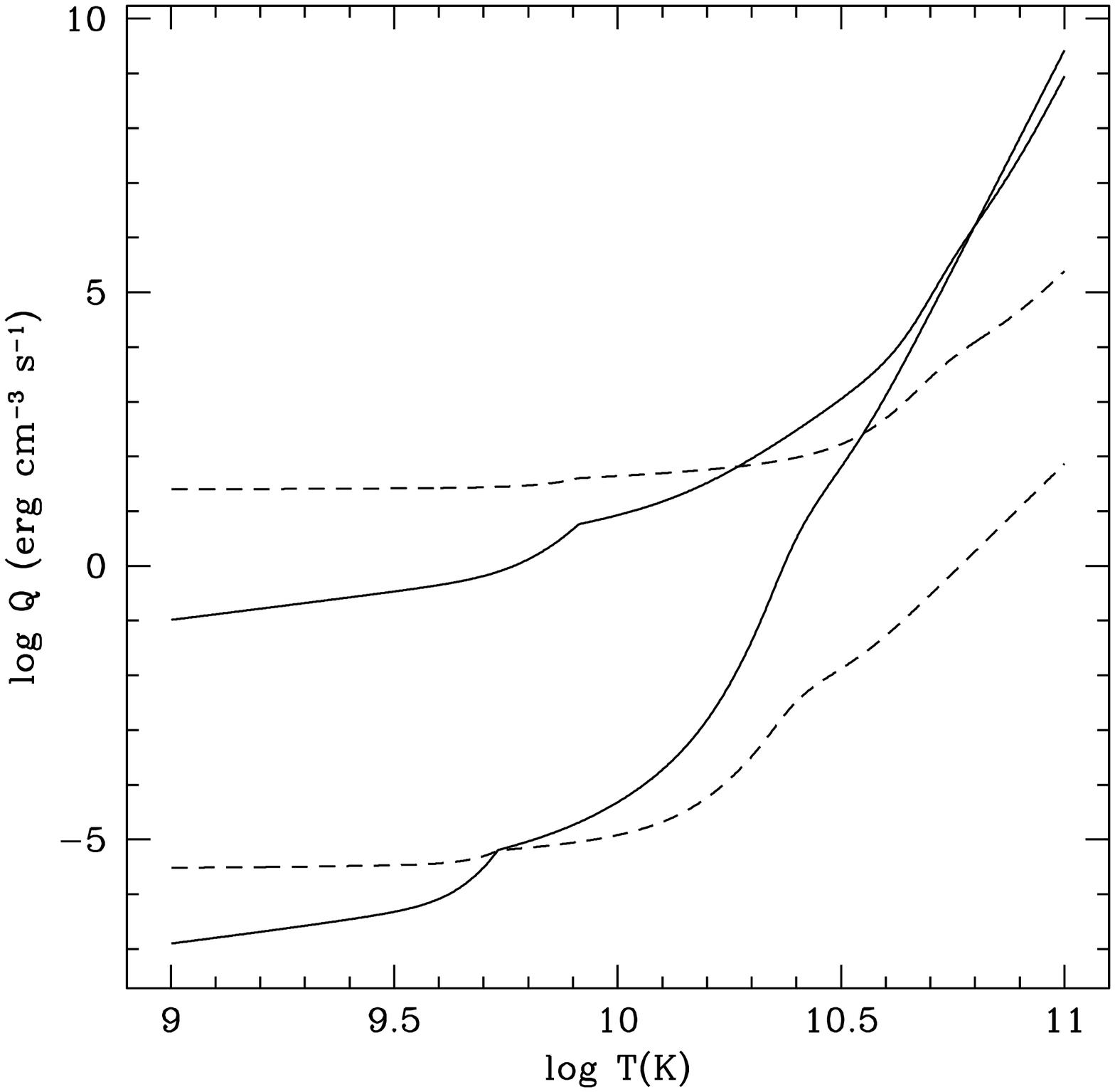}}
\centerline{\epsfysize=4in\epsffile{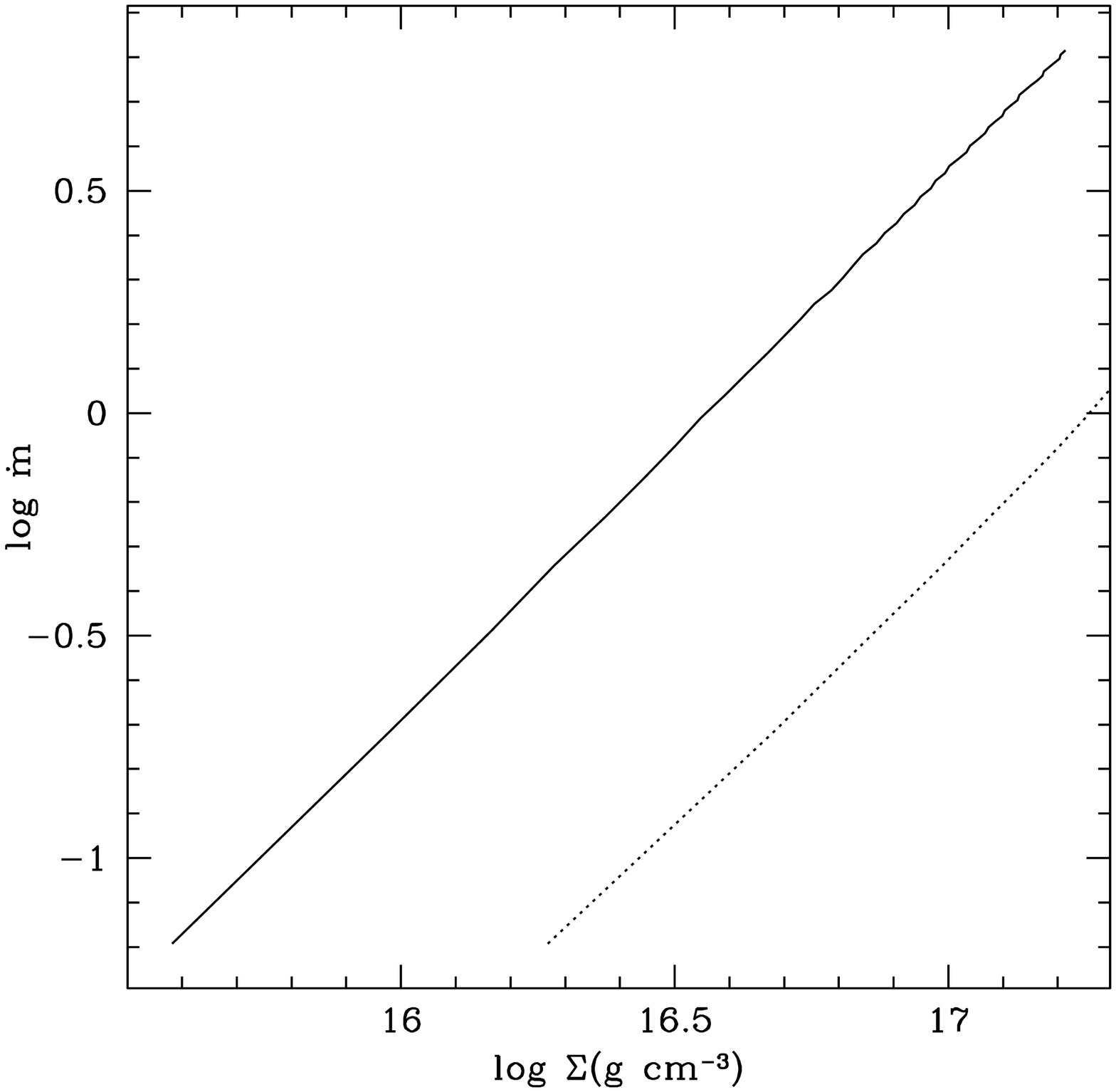}}
\end{figure} 

\begin{figure}[t]
\centerline{\epsfysize=5.7in\epsffile{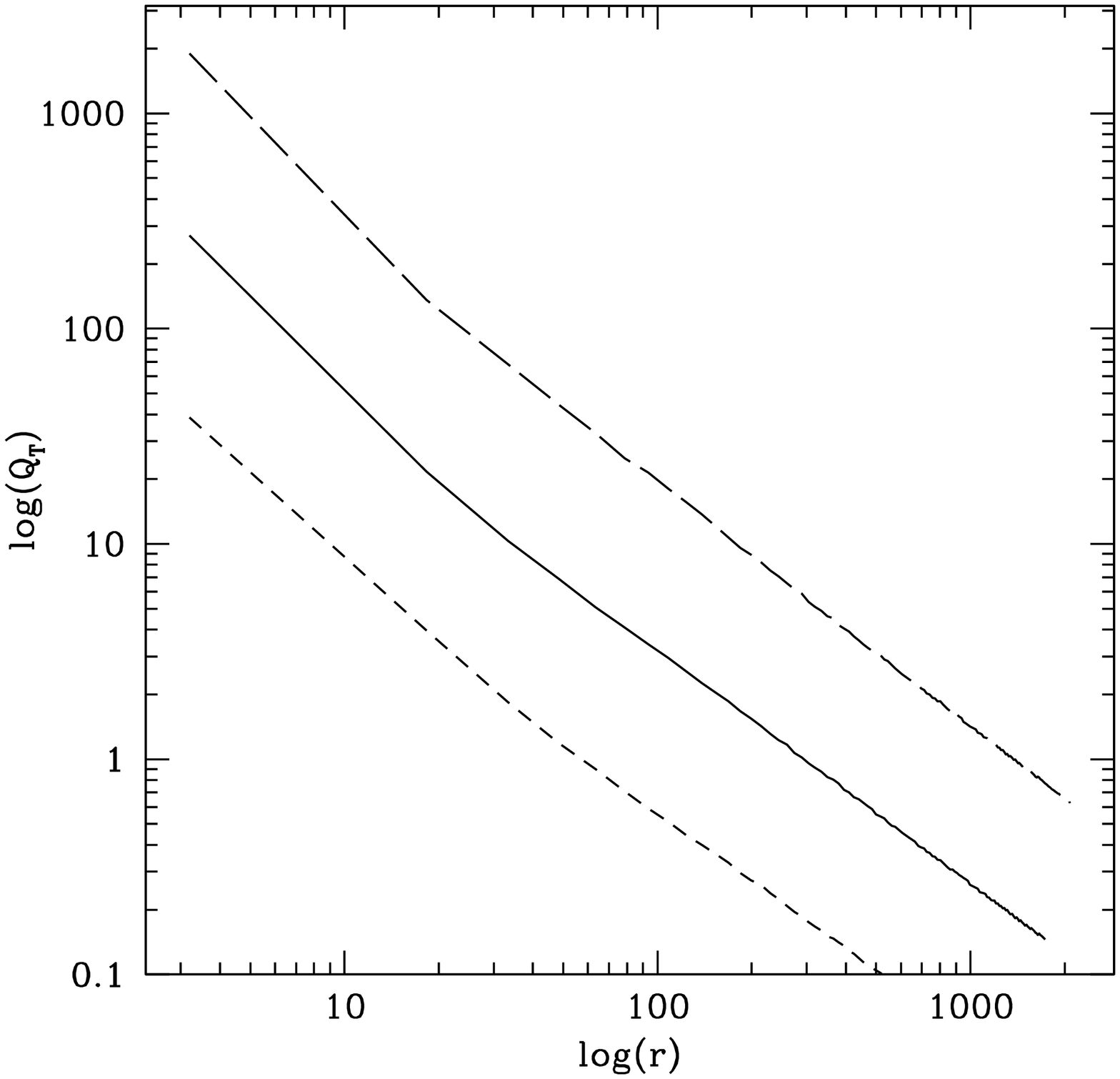}}
\end{figure} 

\begin{figure}[t]
\centerline{\epsfysize=5.7in\epsffile{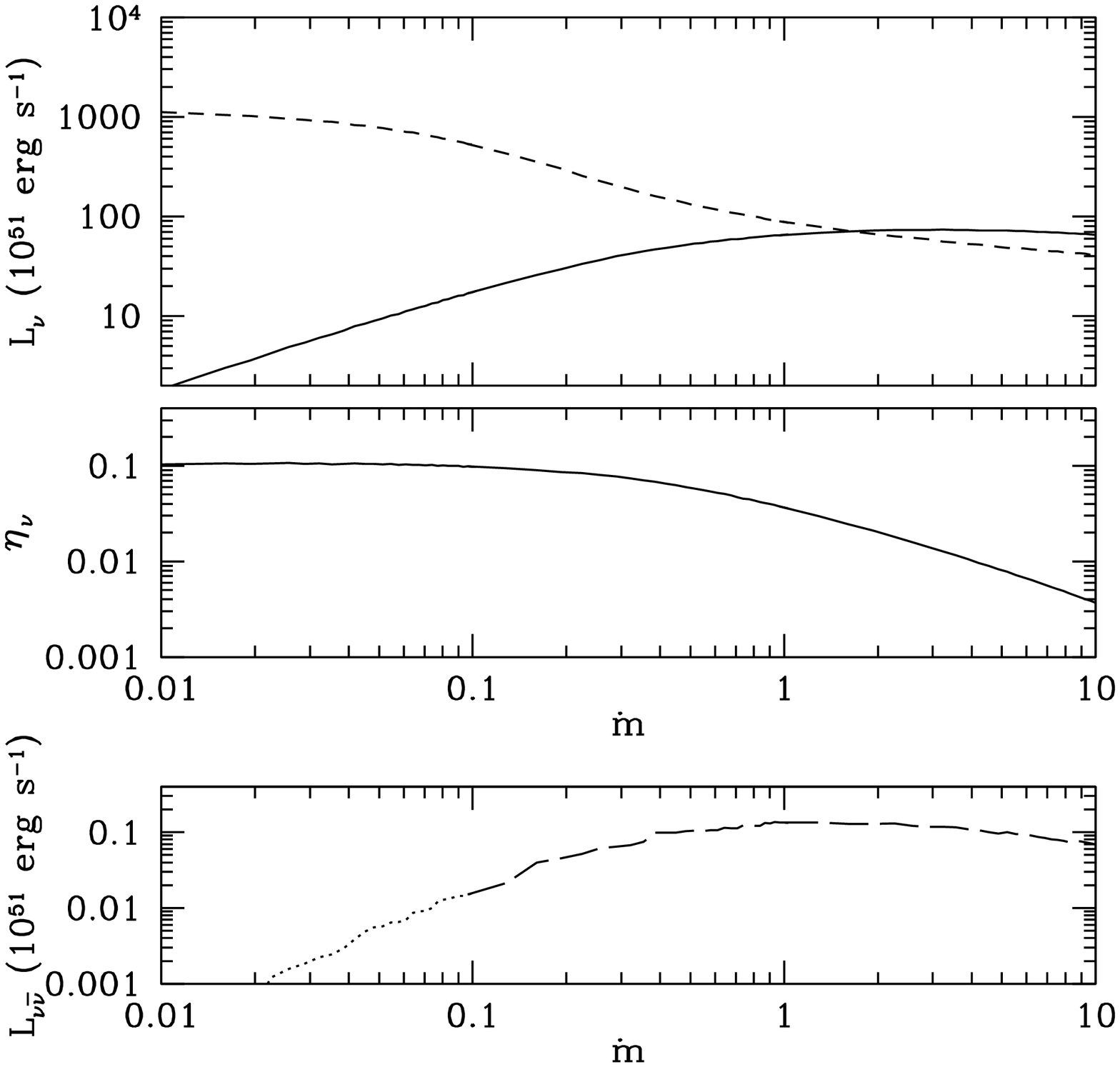}}
\end{figure}

\end{document}